# The Mission Market: Flying NASA's Vetted, Unflown Science

*Author: Dr. Blakesley Burkhart, Associate Professor, Rutgers*

*Summary: Every year NASA validates more excellent science missions than it can afford to fly. Some of the most promising missions in astrophysics, heliophysics, and planetary science are judged selectable, yet remain grounded indefinitely. Mission Market is an advance market commitment that can fly lower cost missions without competing for federal appropriations. Funding comes primarily from the enormous capital now flowing into artificial intelligence (AI) companies that value generating frontier data for scientific foundation models and is also supplemented by traditional means such as philanthropy and patron-scientists.*

NASA's Explorer, Mission of Opportunity, and related programs run a rigorous, multi-year, peer-reviewed competition that produces a standing inventory of fully vetted science mission concepts [1]. These mission concepts range in cost cap from about 50 to 300 million and have produced groundbreaking scientific discovery for a fraction of the cost of Great Observatory class missions (e.g., Hubble and James Webb Space Telescopes), especially given recent reductions in launch cost via commercial rideshare [2][3]. However, only one or two fly per decade. The concepts that lose are, in most cases, not scientifically deficient; they are rated selectable and then left unfunded because the program can only afford to fly a small fraction.

The Mission Market concept proposed here is a policy initiative that applies the advance market commitment (AMC) mechanism, proven for vaccines and carbon removal [4][5], to flight-ready space science. A neutral intermediary, co-chartered with NASA and one or more major science foundations, would commit in advance to fund the flight of vetted missions, escrowing each mission's full cost and disbursing against milestones. The resource barrier has been identified before [6], however what is new is both the mechanism and the data-hungry AI funding landscape that can make it viable now. Frontier scientific observations are among the few data products that cannot be replicated by AI company competitors, scraped from the internet, or synthesized. As scientific foundation models become strategic assets, exclusive early access to newly generated astrophysical, Earth science, heliophysics, and planetary datasets may become valuable enough to justify direct investment in their creation.

The goal is performance-defined: *Fly the highest-rated, cost-effective NASA science missions that peer review has already validated but the federal budget cannot accommodate, funded by industry and private partners.* The program will materially increase the annual flight rate of peer-reviewed space science within a flat federal topline, by mobilizing non-federal capital to fly the small-mission tail NASA cannot afford, and in doing so will seed a competitive commercial market for low-cost science-mission delivery. Critically, the threshold for mission consideration remains the Decadal Surveys and NASA peer review, which preserves scientific quality and keeps mission risk well-understood [7] before any private capital is committed.

## 1. The New Monetary Value of *Scientific Data*

Private and philanthropic funding for space science missions is not a new idea. It has been proposed and attempted before, it remains valid, and it is part of the model here, but on its own it has never reached the scale these missions demand [6]. What is new *now* is the rapid growth of Artificial Intelligence firms and what such firms value. The AI foundation models reshaping the modern economy are only as good as the data they train on, and the high-quality public supply that built today's models is already being exhausted. This is already pushing AI companies to compete for new, original, hard-to-obtain data [8][9]. Observations that exist in no training set are among the most valuable data of all. That change opens a funding stream that did not exist a decade ago, far larger than philanthropy and well suited to *frontier science*. This creates a market connecting that capital to NASA's inventory of unflown, vetted, highly rated science missions. Building it is what an AMC does: commits demand in advance for a product the market has not yet learned to value [4], which is why a grant program would not serve and a market-shaping instrument would.

Corporate participation tied to artificial intelligence is the center of the funding case. The mechanism is to route a small fraction of the capital now flowing into AI toward generating frontier scientific data for the models that increasingly drive discovery. A single publisher data deal can reach a quarter of a billion dollars [9], more than the full cost of a NASA Small Explorer; public and corporate programs are committing nine figure sums to scientific foundation models and the AI-ready datasets they require, including a $152 million investment in open scientific models [8]; and annual AI infrastructure spending now runs to several hundred billion dollars [10]. Against any of these, a complete small NASA space science mission, roughly $50 to $300 million depending on class [1], represents a small fraction of any of those figures and to AI models' training budgets generally. The willingness to pay for data is established and the capital is in place; this initiative directs it at frontier space mission data.

The form of this corporate return is compatible with open science, which is essential because NASA data is public by mandate and must remain so. A funder does not purchase the data but pays for its creation. The return is a head start during the mission's proprietary period before public release, a co-developed benchmark or domain model, a compute and engineering partnership, and the prestige of having enabled a named discovery. Typically NASA-funded missions have a proprietary period of 6 months to 1 year, and similar lengths for the funder would not be unusual. The data itself is released to the community on the normal schedule.

Capital reaches the missions through four tranches. The AI capital leads while the rest are traditional and supporting. Philanthropy provides a base of gifts whose return is impact and identity. Patron-scientists, individuals who fund a mission in return for a genuine role in its science, can broaden that base. Private investors can be a minor, optional source for missions with dual-use value, a route proposed a decade ago [6]. NASA is the keystone across all four tranches: peer-review vetting, the brand and credibility that mobilize trust, engineering partnership through its centers, and fixed-price and risk-tailored flight [7].

## 2. The Market Shaping Approach

Each mission is fully funded to completion at the moment it is authorized. The full cost is committed and escrowed up front, so a mission cannot be half-built and then abandoned. The committed sum is released in stages against verified milestones, from design reviews through integration, launch, and validated data delivery. This is the model NASA used for Commercial Orbital Transportation Services (COTS) and Commercial Lunar Payload Services (CLPS): full fixed price committed at the start, disbursed by milestone, with overruns on the builder [2][3].

A funder engages at the level of a single mission. Because NASA has already vetted every candidate as scientifically sound and held it to a cost cap[1], the funder is not judging quality or steering the science; it is choosing, among proven missions, the one whose data it most wants, e.g., an exoplanet survey rather than a black-hole mission. Funding that mission is what earns the early priority access to its data: an AI firm that expects a mission's observations to matter for its models can fund the flight and be first to use what comes back. The market-shaping effect appears once the commitment covers a stream of missions rather than one. A standing promise to fund many small missions over years is the kind of durable demand that leads commercial vendors to build cheap, standardized spacecraft, the same dynamic by which NASA's CLPS commitment called a commercial lunar industry into being. *Over time that lowers the cost of each mission for every buyer, including NASA, and is what turns a funding instrument into a market [2].*

### 3. Three Reasons Why Mission Market Is Achievable Now

**The demand for frontier scientific data is already established.** AI firms now pay hundreds of millions of dollars for data licensing deals, and public and corporate programs are committing nine-figure sums specifically to build scientific foundation models [8][9][10]. The observations these missions produce, original data that exists in no training set, are among the most sought after of all [11]. Unlike pure philanthropy, which was tried and stalled because the donor pool at the required scale was simply too small [6], this funding stream is driven by commercial self-interest rather than generosity alone. Philanthropic donors and patron-scientists remain welcome participants, and Mission Market lowers the barrier to their involvement by providing a structured, credible vehicle that did not previously exist.

**A government pay-for-success pipeline already created a commercial market:** NASA's Commercial Lunar Payload Services program buys lunar delivery as a fixed-price service, accepts higher per-mission risk for more attempts, and in doing so called a commercial lunar delivery industry into existence [2]. Fixed-price milestone contracting has a proven track record of cutting development costs, the same model that reduced Falcon 9 development cost by 68 percent compared to traditional NASA estimates [3]. The Mission Market applies the same theory of change to small science missions.

**A privately organized AMC provides a successful precedent:** Frontier, founded in 2022, demonstrates that a neutral intermediary can create a functioning market where none existed (in this case, buying carbon removal). It aggregates demand from multiple buyers into an annual spend schedule, runs competitive requests for proposals against published target product profiles, uses an expert technical and commercial review team, and pays through pre-purchases for early stage suppliers and offtake agreements for those ready to scale [4]. It borrowed the structure from the pneumococcal vaccine AMC, which is credited with saving roughly 700,000 lives [5]. This is the operational template for the Mission Market intermediary: binding commitments, competitive mission selection, and milestone disbursement, whether from a single AI anchor or a pool of buyers.

### 4. Guardrails

Four safeguards keep the initiative scientifically sound. First, the private funding channel is confined by design to MIDEX and smaller cost cap missions, while larger cost cap missions (e.g., Probe, Great Observatory class) remain federal, because only sovereign-scale capital can carry their cost, horizon, and risk. Since flagships lie outside the channel's reach, the channel cannot become a rationale for cutting them. Second, scientific merit is protected by drawing only on concepts that have already passed NASA peer review, so funders and patrons choose among vetted missions and never set the science, which remains governed by the Astro Decadal Surveys. Third, open science is protected by mandating normal public release of all data, with private value limited to a bounded proprietary window. Finally, mission assurance is protected by flying in partnership with NASA and its centers and by managing risk across a portfolio, rather than by pretending private capital buys equivalent reliability for less. The NASA selection-to-flight process includes significant margins on cost, which provides additional overrun guardrails.

### 5. Measuring Success and Timeline

- Additional selectable missions flown per year, versus the program baseline.
- Realized cost per mission, versus the program cost cap.
- Concept-to-launch time reduced from the current federal only model.
- Science output: validated data products, peer-reviewed results, and AI-ready datasets released.
- Net new commercial providers entering the science-mission-as-a-service market.

- **Phase 0 (Year 0 to 1):** Charter the intermediary; secure anchor commitments from a foundation, a NASA letter of intent, and a corporate AI anchor; publish the first target product profiles. A seed AMC in the low hundreds of millions over the first window is sufficient to begin. Additional open questions for the program, e.g. regarding smaller costs such as launch costs, will be addressed.
- **Phase 1, pilot (Years 1 to 4):** Fly two to three Mission-of-Opportunity and Smallsat class missions, each under roughly 60 million dollars and some well below, using rideshare, commercial buses, and a Class-D risk posture.
- **Phase 2, scale (Years 4 to 8):** Operate a standing annual pipeline across the Mission-of-Opportunity and SMEX tiers and demonstrate commercial-provider market formation.